\documentclass[10pt,a4paper,onecolumn]{IEEEtran}
\usepackage[lmargin=2cm, rmargin=2cm, tmargin=2cm, bmargin=2cm]{geometry}
\usepackage{graphicx}
\usepackage{mathptmx}
\usepackage{times}
\usepackage{colortbl}
\usepackage[authoryear,round]{natbib}
\usepackage[T1]{fontenc}
\usepackage[latin1]{inputenc}
\usepackage{subfigure}
\usepackage{amsmath}
\usepackage{amsfonts}
\usepackage{amssymb}
\newtheorem{assumption}{Assumption}
\newtheorem{remark}{Remark}

\DeclareMathOperator{\sgn}{sgn}
\DeclareMathOperator{\sat}{sat}

\title{Sliding mode control with a neural network compensation scheme for electro-hydraulic systems}

\author{Josiane Maria de Macedo Fernandes, Marcelo Costa Tanaka, Wallace Moreira Bessa}

\date{}

\begin{document}

\maketitle

\abstract{

Electro-hydraulic servo-systems are widely employed in industrial applications such as robotic manipulators, active suspensions, 
precision machine tools and aerospace systems. They provide many advantages over electric motors, including high force to weight 
ratio, fast response time and compact size. However, precise control of electro-hydraulic systems, due to their inherent nonlinear 
characteristics, cannot be easily obtained with conventional linear controllers. Most flow control valves can also exhibit some 
hard nonlinearities such as dead-zone due to valve spool overlap. This work describes the development of a sliding mode controller
with a neural network compensation scheme for electro-hydraulic systems subject to an unknown dead-zone input. The boundedness and 
convergence properties of the closed-loop signals are proven using Lyapunov stability theory. Numerical results are presented in 
order to demonstrate the control system performance.

}

\section*{INTRODUCTION}

Electro-hydraulic actuators play an essential role in several branches of industrial activity and are frequently the most suitable 
choice for systems that require large forces at high speeds. Their application scope ranges from robotic manipulators to aerospace 
systems. Another great advantage of hydraulic systems is the ability to keep up the load capacity, which in the case of electric 
actuators is limited due to excessive heat generation. 

However, the dynamic behavior of electro-hydraulic systems is highly nonlinear, which in fact makes the design of controllers for 
such systems a challenge for the conventional and well established linear control methodologies. The increasing number of works 
dealing with control approaches based on modern techniques shows the great interest of the engineering community, both in academia 
and industry, in this particular field. The most common approaches are the adaptive \citep{guan1,guan2,yanada1,yao1} and variable 
structure \citep{jirs2010,mihajlov1,bonchis1,liu1} methodologies, but nonlinear controllers based on quantitative feedback theory 
\citep{niksefat1}, optimal tuning PID \citep{liu2}, adaptive neural network \citep{knohl1} and adaptive fuzzy system 
\citep{jbsmse2010} were also presented.

In addition to the common nonlinearities that originate from the compressibility of the hydraulic fluid and valve flow-pressure 
properties, most electro-hydraulic systems are also subjected to hard nonlinearities such as dead-zone due to valve spool overlap. 
It is well-known that the presence of a dead-zone can lead to performance degradation of the controller and limit cycles or even 
instability in the closed-loop system. To overcome the negative effects of the dead-zone nonlinearity, many works 
\citep{tao1,kim1,oh1,selmic1,tsai1,zhou1} use an inverse function even though this approach leads to a discontinuous control law 
and requires instantaneous switching, which in practice can not be accomplished with mechanical actuators. An alternative scheme, 
without using the dead-zone inverse, was originally proposed by \citet{lewis1} and also adopted by \citet{wang1}. In both works, 
the dead-zone is treated as a combination of a linear and a saturation function. This approach was further extended by 
\citet{ibrir1} and \citet{zhang1}, in order to accommodate non-symmetric and unknown dead-zones, respectively.

Intelligent control, on the other hand, has proven to be a very attractive approach to cope with uncertain nonlinear systems 
\citep{tese,rsba2010,cobem2005,nd2012,Bessa2014,Bessa2017,Bessa2018,Bessa2019,Deodato2019,Lima2018,Lima2020,Lima2021,Tanaka2013}. 
By combining nonlinear control techniques, such as feedback linearization or sliding modes, with adaptive intelligent algorithms, 
for example fuzzy logic or artificial neural networks, the resulting intelligent control strategies can deal with the nonlinear 
characteristics as well as with modeling imprecisions and external disturbances that can arise.

In this work, a sliding mode controller with a neural network compensation scheme is proposed for electro-hydraulic systems subject 
to an unknown dead-zone input. The adopted approach does not require previous knowledge of dead-zone parameters nor the construction 
of an inverse function. On this basis, a smooth sliding mode controller is considered to confer robustness against modeling 
imprecisions and a Radial Basis Function (RBF) neural network is embedded in the boundary layer to cope with dead-zone effects. 
The boundedness and convergence properties of the closed-loop system are analytically proven using Lyapunov stability theory. 
Numerical simulations are carried out in order to demonstrate the control system performance.

\section*{ELECTRO-HYDRAULIC SYSTEM MODEL}

In order to design the adaptive fuzzy controller, a mathematical model that represents the hydraulic 
system dynamics is needed. Dynamic models for such systems are well documented in the literature
\citep{merritt1,walters1}.

\begin{figure}[htb]
\centering
\includegraphics[width=0.55\textwidth]{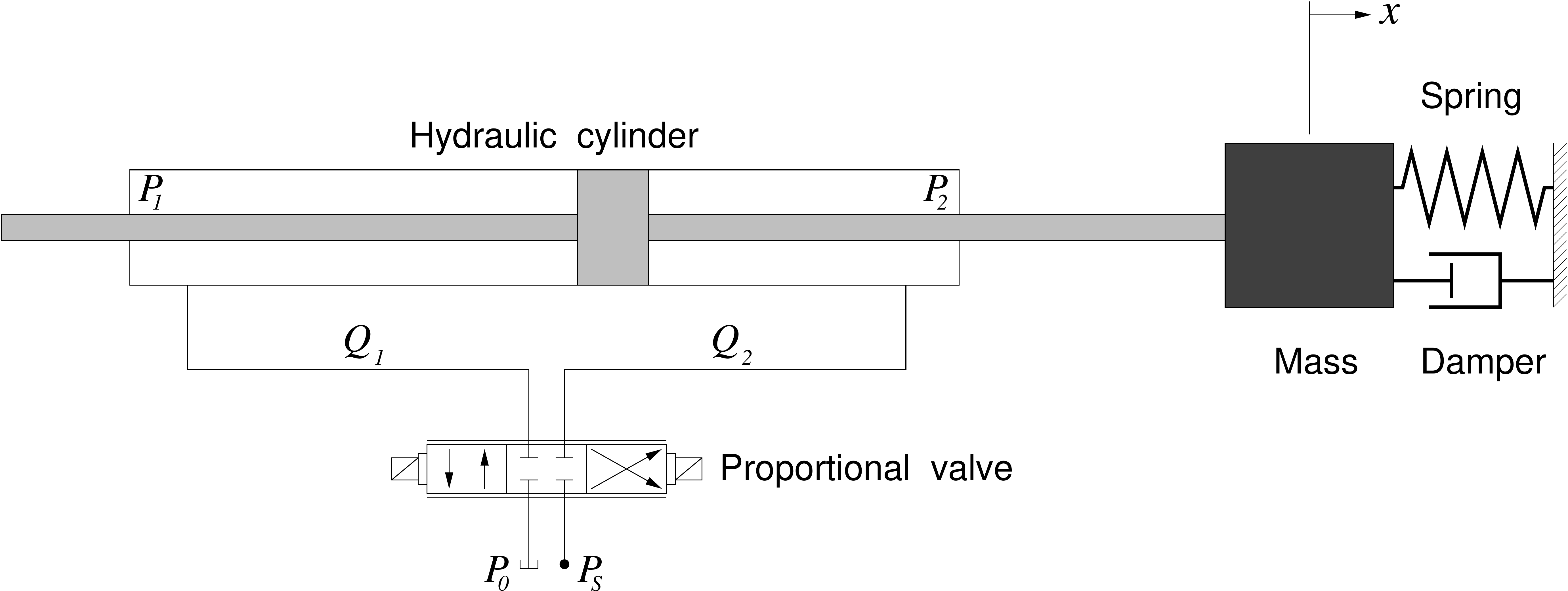}
\caption{Schematic diagram of the electro-hydraulic servo-system.}
\label{fi:sistema}
\end{figure}

The electro-hydraulic system considered in this work consists of a four-way proportional valve, 
a hydraulic cylinder and variable load force. The variable load force is represented by a 
mass--spring--damper system. The schematic diagram of the system under study is presented in
Fig.~\ref{fi:sistema}.

The balance of forces on the piston leads to the following equation of motion: 

\begin{equation}
F_g=A_1P_1-A_2P_2=M_t\ddot{x}+B_t\dot{x}+K_s{x}
\label{eq:mov_aux}
\end{equation}

\noindent
where $F_g$ is the force generated by the piston, $P_1$ and $P_2$ are the pressures at each side of 
cylinder chamber, $A_1$ and $A_2$ are the ram areas of the two chambers, $M_t$ is the total mass of
piston and load referred to piston, $B_t$ is the viscous damping coefficient of piston and load, 
$K_s$ is the load spring constant and $x$ is the piston displacement.

Defining the pressure drop across the load as $P_l=P_1-P_2$ and considering that for a symmetrical
cylinder $A_p=A_1=A_2$, Eq.~(\ref{eq:mov_aux}) can be rewritten as

\begin{equation}
M_t\ddot{x}+B_t\dot{x}+K_s{x}=A_pP_l
\label{eq:mov}
\end{equation}

Applying continuity equation to the fluid flow, the following equation is obtained:

\begin{equation}
Q_l=A_p\dot{x}+C_{tp}+\frac{V_t}{4\beta_e}\dot{P}_l
\label{eq:cont}
\end{equation}

\noindent
where $Q_l=(Q_1+Q_2)/2$ is the load flow, $C_{tp}$ the total leakage coefficient of piston, $V_t$ the 
total volume under compression in both chambers and $\beta_e$ the effective bulk modulus.

Considering that the return line pressure is usually much smaller than the other pressures involved
($P_0\approx0$) and assuming a closed center spool valve with matched and symmetrical orifices, the 
relationship between load pressure $P_l$ and load flow $Q_l$ can be described as follows

\begin{equation}
Q_l=C_dw\bar{x}_{sp}\sqrt{\frac{1}{\rho}\big(P_s-\sgn(\bar{x}_{sp})P_l\big)}
\label{eq:fluxo}
\end{equation}

\noindent
where $C_d$ is the discharge coefficient, $w$ the valve orifice area gradient, $\bar{x}_{sp}$ the 
effective spool displacement from neutral, $\rho$ the hydraulic fluid density, $P_s$ the supply 
pressure and $\sgn(\cdot)$ is defined by

\begin{displaymath}
\sgn(z) = \left\{\begin{array}{rc}
-1&\mbox{if}\quad z<0 \\
0&\mbox{if}\quad z=0 \\
1&\mbox{if}\quad z>0
\end{array}\right.
\end{displaymath}

\begin{figure}[hbt]
\centering
\includegraphics[width=0.3\textwidth]{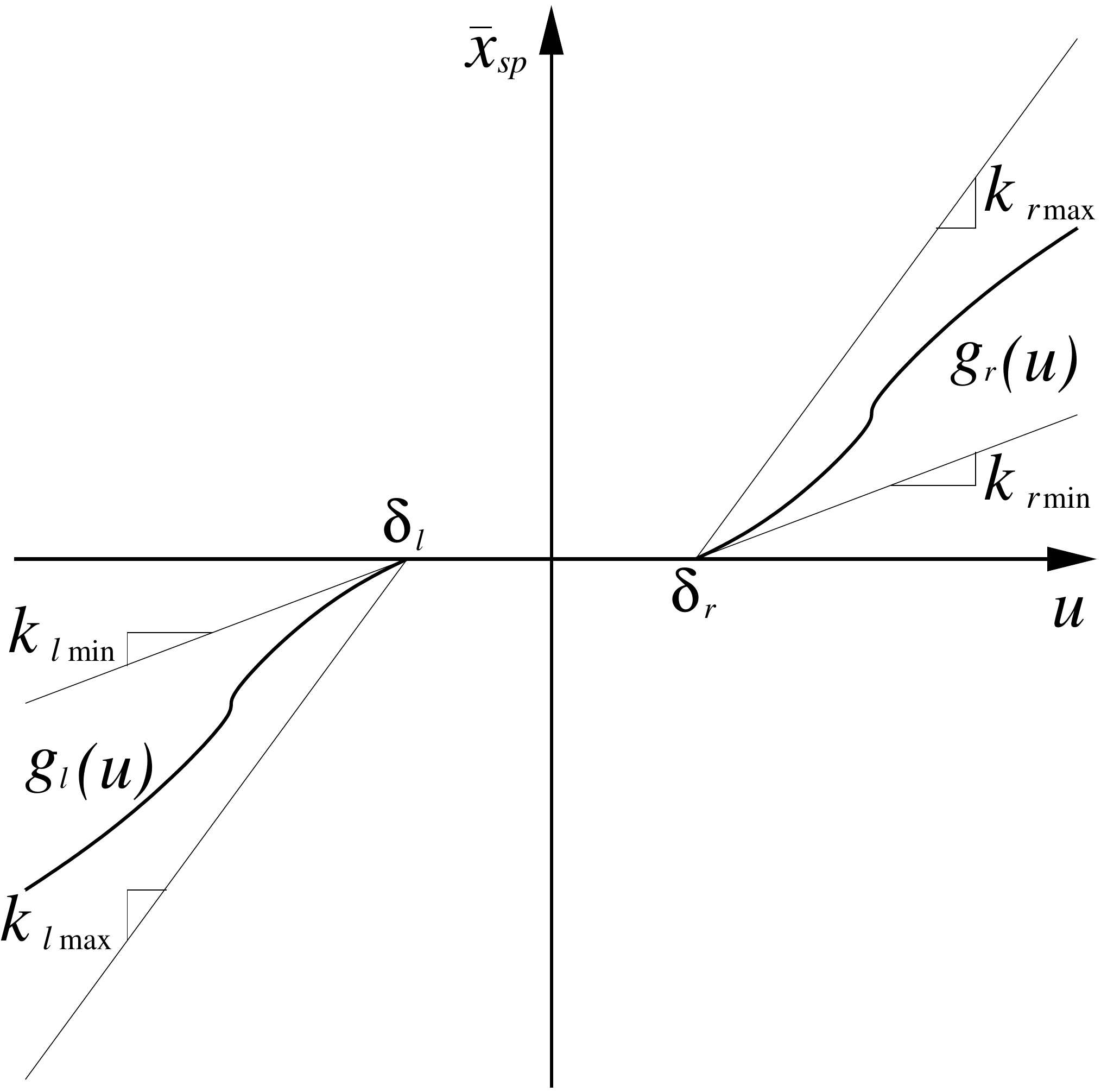} 
\caption{Dead-zone nonlinearity.}
\label{fi:dzone}
\end{figure}

Assuming that the dynamics of the valve are fast enough to be neglected, the valve spool displacement 
can be considered as proportional to the control voltage ($u$). For closed center valves, or even in the 
case of the so-called critical valves, the spool presents some overlap. This overlap prevents from leakage 
losses but leads to a dead-zone nonlinearity within the control voltage, as shown in Fig.~\ref{fi:dzone}.

The adopted dead-zone model is a slightly modified version of that proposed by \citet{zhang1}, 
which can be mathematically described by 

\begin{equation}
\bar{x}_{sp}= \left\{\begin{array}{ll}
g_l(u)&\mbox{if}\quad u\le\delta_l\\
0&\mbox{if}\quad  \delta_l<u<\delta_r\\
g_r(u)&\mbox{if}\quad u\ge\delta_r 
\end{array}\right.
\label{eq:dzone1}
\end{equation}

\noindent
where $g_l$ and $g_r$ are functions of control voltage and the dead-band parameters $\delta_l$ and 
$\delta_r$ depends on the size of the overlap region.

In respect of the dead-zone model presented in Eq.~(\ref{eq:dzone1}), the following assumptions can 
be made: 

\begin{assumption}
The dead-zone output $\bar{x}_{sp}$ is not available to be measured.
\label{as:output}
\end{assumption}
\begin{assumption}
The dead-band parameters $\delta_l$ and $\delta_r$ are unknown but bounded and with known signs, 
i.e., $\delta_{l\,\mathrm{min}}\le\delta_l\le\delta_{l\,\mathrm{max}}<0$ and $0<\delta_{r\,
\mathrm{min}}\le\delta_r\le\delta_{r\,\mathrm{max}}$.
\label{as:dband}
\end{assumption}
\begin{assumption}
The functions $g_l:(-\infty,\delta_l]$ and $g_r:[\delta_r,+\infty)$ are $C^1$ and with bounded 
positive-valued derivatives, i.e., 
\begin{displaymath}
0<k_{l\,\mathrm{min}}\le g'_l(u)\le k_{l\,\mathrm{max}},\quad\forall u\in(-\infty,\delta_l],
\end{displaymath}
\begin{displaymath}
0<k_{r\,\mathrm{min}}\le g'_r(u)\le k_{r\,\mathrm{max}},\quad\forall u\in[\delta_r,+\infty),
\end{displaymath}

\noindent
where $g'_l(u)=dg_l(z)/dz|_{z=u}$ and $g'_r(u)=dg_r(z)/dz|_{z=u}$.
\label{as:slopes}
\end{assumption}

\begin{remark}
Assumption~\ref{as:slopes} means that both $g_l$ and $g_r$ are Lipschitz functions. 
\end{remark}

From the mean value theorem and noting that $g_l(\delta_l)=g_r(\delta_r)=0$, it follows that there 
exist $\xi_l:\mathbb{R}\to(-\infty,\delta_l)$ and $\xi_r:\mathbb{R}\to(\delta_r,+\infty)$ such 
that

\begin{displaymath}
g_l(u)=g'_l\big(\xi_l(u)\big)[u-\delta_l]
\end{displaymath}
\begin{displaymath}
g_r(u)=g'_r\big(\xi_r(u)\big)[u-\delta_r]
\end{displaymath}

In this way, Eq.~(\ref{eq:dzone1}) can be rewritten as follows:

\begin{equation}
\bar{x}_{sp} = \left\{\begin{array}{ll}
g'_l\big(\xi_l(u)\big)[u-\delta_l]&\mbox{if}\quad u\le\delta_l\\
0&\mbox{if}\quad  \delta_l<u<\delta_r\\
g'_r\big(\xi_r(u)\big)[u-\delta_r]&\mbox{if}\quad u\ge\delta_r 
\end{array}\right.
\label{eq:dzone2}
\end{equation}

\noindent
or in a more appropriate form: 

\begin{equation}
\bar{x}_{sp}=k_v(u)[u-d(u)] 
\label{eq:dzone3}
\end{equation}

\noindent
where

\begin{equation}
k_v(u)= \left\{\begin{array}{ll}
g'_l\big(\xi_l(u)\big)&\mbox{if}\quad u\le0\\
g'_r\big(\xi_r(u)\big)&\mbox{if}\quad u>0 
\end{array}\right.
\label{eq:slopes}
\end{equation}
\quad \mbox{ and } \quad
\noindent
and

\begin{equation}
d(u)= \left\{\begin{array}{ll}
\delta_l&\mbox{if}\quad u\le\delta_l\\
u&\mbox{if}\quad  \delta_l<u<\delta_r\\
\delta_r&\mbox{if}\quad u\ge\delta_r 
\end{array}\right.
\label{eq:dsat}
\end{equation}

\begin{remark}
Considering Assumption~\ref{as:dband} and Eq.~(\ref{eq:dsat}), it can be easily verified that $d(u)$ 
is bounded: $|d(u)|\le\delta$, where $\delta=\mathrm{max}\{-\delta_{l\,\mathrm{min}},\delta_{r\,
\mathrm{max}}\}$.
\end{remark}

Combining equations (\ref{eq:mov}), (\ref{eq:cont}), (\ref{eq:fluxo}) and (\ref{eq:dzone3}) leads 
to a third-order differential equation that represents the dynamic behavior of the electro-hydraulic 
system:

\begin{equation}
\dddot{x}=-\mathbf{a^\mathrm{T}x}+b(\mathbf{x},u)u-b(\mathbf{x},u)d(u)
\label{eq:system}
\end{equation}

\noindent
where $\mathbf{x}=[x,\dot{x},\ddot{x}]$ is the state vector with an associated coefficient vector 
$\mathbf{a}=[a_0,a_1,a_2]$ defined according to

\begin{displaymath}
\displaystyle
a_0=\frac{4\beta_eC_{tp}K_s}{V_tM_t}\quad;\quad
a_1=\frac{K_s}{M_t}+\frac{4\beta_eA_{p}^2}{V_tM_t}+\frac{4\beta_eC_{tp}B_t}{V_tM_t}\quad;\quad
a_2=\frac{B_t}{M_t}+\frac{4\beta_eC_{tp}}{V_t}
\end{displaymath}

\noindent
and

\begin{displaymath}
\displaystyle
b(\mathbf{x},u)=\frac{4\beta_eA_p}{V_tM_t}C_dwk_v\sqrt{\frac{1}{\rho}\big[P_s-\sgn(u)
\big(M_t\ddot{x}+B_t\dot{x}+K_s{x}\big)/A_p\big]}
\end{displaymath}

In respect of the dynamic system presented in Eq.~(\ref{eq:system}), the following assumptions will 
also be made:

\begin{assumption}
The coefficients $a_0$, $a_1$ and $a_2$ are unknown but bounded: $|(\mathbf{\hat{a}-a})^\mathrm{T}
\mathbf{x}|\le\alpha$, where $\mathbf{\hat{a}}$ is an estimate of $\mathbf{a}$.
\label{as:abounds}
\end{assumption}
\begin{assumption}
The input gain $b(\mathbf{x},u)$ is unknown but positive and bounded: $0<b_{\mathrm{min}}\le
b(\mathbf{x},u)\le b_{\mathrm{max}}$.
\label{as:bbounds}
\end{assumption}

Based on the dynamic model presented in Eq.~(\ref{eq:system}), a neural network sliding mode controller is developed.

\section*{NEURAL NETWORK BASED SLIDING MODE CONTROLLER}

The proposed control problem is to ensure that, even in the presence of parametric uncertainties, 
unmodeled dynamics and an unknown dead-zone input, the state vector $\mathbf{x}$ will follow a desired 
trajectory $\mathbf{x}_d=[x_d,\dot{x}_d,\ddot{x}_d]$ in the state space.

Regarding the development of the control law, the following assumptions should also be made:

\begin{assumption}
The state vector $\mathbf{x}$ is available.
\label{as:stat}
\end{assumption}
\begin{assumption}
The desired trajectory $\mathbf{x}_d$ is once differentiable in time. Furthermore, every element
of vector $\mathbf{x}_d$, as well as $\dddot{x}_d$, is available and with known bounds.
\label{as:traj}
\end{assumption}

Let $\tilde{x}=x-x_d$ be defined as the tracking error in the variable $x$, $\mathbf{\tilde{x}}=
\mathbf{x}-\mathbf{x}_d=[\tilde{x},\dot{\tilde{x}},\ddot{\tilde{x}}]$ as the tracking error vector 
and consider a sliding surface $S$ defined in the state space by the equation $s(\mathbf{\tilde{x}})
=0$, with the function $s:\mathbb{R}^3\to\mathbb{R}$ satisfying 

\begin{equation}
s(\mathbf{\tilde{x}})=\ddot{\tilde{x}}+2\lambda\dot{\tilde{x}}+\lambda^2\tilde{x}
\label{eq:surf}
\end{equation}

\noindent 
where $\lambda$ is a strictly positive constant.

Now, the problem of controlling the system dynamics (\ref{eq:system}) can be treated according to 
the sliding mode methodology, by defining a control law composed by an equivalent control $\hat{u}
=\hat{b}^{-1}(\mathbf{\hat{a}^\mathrm{T}x}+\dddot{x}_d-2\lambda\ddot{\tilde{x}}-\lambda^2
\dot{\tilde{x}})$, an estimate $\hat{d}$ and a discontinuous term $-K\sgn(s)$:

\begin{equation}
u=\hat{b}^{-1}(\mathbf{\hat{a}^\mathrm{T}x}+\dddot{x}_d-2\lambda\ddot{\tilde{x}}-\lambda^2
\dot{\tilde{x}})+\hat{d}-K\sgn(s)
\label{eq:usgn}
\end{equation}

\noindent
where $K$ is the control gain.

Based on Assumption~\ref{as:bbounds} and considering that the estimate $\hat{b}$ could be chosen
according to the geometric mean $\hat{b}=\sqrt{b_\mathrm{max}b_\mathrm{min}}$, the bounds of $b$ may 
be expressed as $\gamma^{-1}\le\hat{b}/b\le\gamma$, where $\gamma=\sqrt{b_\mathrm{max}/b_\mathrm{min}}$.

Under this condition, the gain $K$ should be chosen according to:

\begin{equation}
K\ge\gamma\hat{b}^{-1}(\eta+\alpha)+\delta+|\hat{d}|+(\gamma-1)|\hat{u}|
\label{eq:gain}
\end{equation}

\noindent
where $\eta$ is a strictly positive constant related to the reaching time.

At this point, it should be highlighted that the control law (\ref{eq:usgn}), together with 
(\ref{eq:gain}), is sufficient to impose the sliding condition 

\begin{displaymath}
\displaystyle
\frac{1}{2}\frac{d}{dt}s^2\le-\eta|s|
\end{displaymath}

\noindent
and, consequently, the finite time convergence to the sliding surface $S$.

In spite of the demonstrated properties of the controller, the presence of a discontinuous term in
the control law leads to the well known chattering phenomenon. In order to overcome the undesirable 
chattering effects, a thin boundary layer, $S_\phi$, in the neighborhood of the switching surface 
can be adopted \citep{slotine2}:

\begin{displaymath}
\displaystyle
S_\phi=\big\{\mathbf{\tilde{x}}\in\mathbb{R}^3\:\big|\:|s(\mathbf{\tilde{x}})|\le\phi\big\}
\end{displaymath}

\noindent
where $\phi$ is a strictly positive constant that represents the boundary layer thickness.

The boundary layer is achieved by replacing the sign function by a continuous interpolation inside
$S_\phi$. It should be noted that this smooth approximation must behave exactly like the sign
function outside the boundary layer. There are several options to smooth out the ideal relay but
the most common choice is the saturation function:

\begin{displaymath}
\displaystyle
\mbox{sat}(s/\phi) = \left\{\begin{array}{cc}
\mbox{sgn}(s)&\mbox{if}\quad |s/\phi|\ge1 \\
s/\phi&\mbox{if}\quad |s/\phi|<1 
\end{array}\right.
\end{displaymath}

In this way, to avoid chattering, a smooth version of Eq.~(\ref{eq:usgn}) is defined:

\begin{equation}
u=\hat{b}^{-1}(\mathbf{\hat{a}^\mathrm{T}x}+\dddot{x}_d-2\lambda\ddot{\tilde{x}}-\lambda^2
\dot{\tilde{x}})+\hat{d}-K\sat(s/\phi) 
\label{eq:usat}
\end{equation}

In order to establish the attractiveness and invariant properties of the defined boundary layer,
let a new Lyapunov function candidate $V$ be defined as

\begin{displaymath}
\displaystyle
V(t)=\frac{1}{2}s^2_\phi
\end{displaymath}

\noindent
where $s_\phi$ is a measure of the distance of the current state to the boundary layer, and can be
computed as follows

\begin{equation}
s_\phi=s-\phi\:\sat(s/\phi)
\label{eq:dist}
\end{equation}

Noting that $s_\phi=0$ inside the boundary layer and $\dot{s}_\phi=\dot{s}$, one has $\dot{V}(t)=0$ 
inside $S_\phi$, and outside

\begin{displaymath}
\dot{V}(t)=s_\phi\dot{s}_\phi=s_\phi\dot{s}
= (\dddot{\tilde{x}}+2\lambda\ddot{\tilde{x}}+\lambda^2\dot{\tilde{x}})s_\phi
=(-\mathbf{a^\mathrm{T}x}+bu-bd-\dddot{x}_d+2\lambda\ddot{\tilde{x}}+\lambda^2\dot{\tilde{x}})s_\phi
\end{displaymath}

It can be easily verified that outside the boundary layer the control law (\ref{eq:usat}) takes the
following form:

\begin{displaymath}
u=\hat{b}^{-1}(\mathbf{\hat{a}^\mathrm{T}x}+\dddot{x}_d-2\lambda\ddot{\tilde{x}}-\lambda^2
\dot{\tilde{x}})+\hat{d}-K\sgn(s_\phi) 
\end{displaymath}

Thus, the time derivative $\dot{V}$ can be written as

\begin{displaymath}
\dot{V}(t)=-[bK\sgn(s_\phi)-(\mathbf{\hat{a}-a})^\mathrm{T}\mathbf{x}+\hat{b}\hat{u}-b\hat{u}
-b\hat{d}+bd]s_\phi
\end{displaymath}

Therefore, by considering Assumptions~\ref{as:dband}--\ref{as:bbounds} and defining $K$ according to 
(\ref{eq:gain}), $\dot{V}(t)$ becomes:

\begin{equation}
\dot{V}(t)\le-\eta|s_\phi|
\label{eq:liap2d}
\end{equation}

\noindent
which implies $V(t)\le V(0)$ and that $s_\phi$ is bounded. The definitions of $s$ and $s_\phi$, 
respectively Eqs.~(\ref{eq:surf})~and~(\ref{eq:dist}), imply that $\mathbf{\tilde{x}}$ is bounded. 
From the definition of $\dot{s}$ and Assumption~\ref{as:traj} it can be verified that $\dot{s}$ is 
also bounded.

The finite-time convergence of the states to the boundary layer can be shown by integrating both sides
of (\ref{eq:liap2d}) over the interval $0\le t\le t_\mathrm{reach}$, where $t_\mathrm{reach}$ is the
time required to hit $S_\phi$. In this way, noting that $|s_\phi(t_\mathrm{reach})|=0$, one
has:

\begin{equation}
t_\mathrm{reach}\le\frac{|s_\phi(0)|}{\eta}
\label{eq:treach}
\end{equation}

\noindent
which guarantees the convergence of the tracking error vector to the boundary layer in a time
interval smaller than $|s_\phi(0)|/\eta$.

Nevertheless, it should be emphasized that the substitution of the discontinuous term by a smooth
approximation inside the boundary layer turns the perfect tracking into a tracking with guaranteed
precision problem, which actually means that a steady-state error will always remain. However, it can
be verified that, once inside the boundary layer, the tracking error vector will exponentially
converge to a closed region $\Phi$.

Considering that $|s(\mathbf{\tilde{x}})|\le\phi$ may be rewritten as $-\phi\le s(\mathbf{\tilde{x}})
\le\phi$ and from the definition of $s(\mathbf{\tilde{x}})$, Eq.~(\ref{eq:surf}), one has:

\begin{equation}
-\phi\le\ddot{\tilde{x}}+2\lambda\dot{\tilde{x}}+\lambda^2\tilde{x}\le\phi
\label{eq:sbounds}
\end{equation}

Multiplying (\ref{eq:sbounds}) by $e^{\lambda t}$ and integrating between $0$ and $t$:

\begin{gather*}
-\phi e^{\lambda t}\le(\ddot{\tilde{x}}+2\lambda\dot{\tilde{x}}+\lambda^2\tilde{x})e^{\lambda t}
\le\phi e^{\lambda t}\\
-\phi e^{\lambda t}\le\frac{d^{\:2}}{dt^{\:2}}(\tilde{x}e^{\lambda t})\le\phi e^{\lambda t}\\
-\phi\int_0^te^{\lambda\tau}d\tau\le\int_0^t\frac{d^{\:2}}{d\tau^{\:2}}(\tilde{x}e^{\lambda\tau})
d\tau\le\phi\int_0^te^{\lambda\tau}d\tau\\
-\frac{\phi}{\lambda}e^{\lambda t}+\frac{\phi}{\lambda}\le\frac{d}{dt}(\tilde{x}e^{\lambda t})
-\left.\frac{d}{dt}(\tilde{x}e^{\lambda t})\right|_{t=0}\le\frac{\phi}{\lambda}e^{\lambda t}
-\frac{\phi}{\lambda}
\end{gather*}

\noindent
or conveniently rewritten as

\begin{equation}
-\frac{\phi}{\lambda}e^{\lambda t}-\left(\left|\frac{d}{dt}(\tilde{x}e^{\lambda t})\right|_{t=0}+
\frac{\phi}{\lambda}\right)\le\frac{d}{dt}(\tilde{x}e^{\lambda t})\le\frac{\phi}{\lambda}e^{\lambda
t}+\left(\left|\frac{d}{dt}(\tilde{x}e^{\lambda t})\right|_{t=0}+\frac{\phi}{\lambda}\right)
\label{eq:int1bounds}
\end{equation}

Now, integrating (\ref{eq:int1bounds}) between $0$ and $t$:

\begin{equation}
-\frac{\phi}{\lambda^2}e^{\lambda t}-\left(\left|\frac{d}{dt}(\tilde{x}e^{\lambda t})\right|_{t=0}+
\frac{\phi}{\lambda}\right)t-\left(|\tilde{x}(0)|+\frac{\phi}{\lambda^2}\right)\le\tilde{x}e^{\lambda
t}\le\frac{\phi}{\lambda^2}e^{\lambda t}\\+\left(\left|\frac{d}{dt}(\tilde{x}e^{\lambda t})\right|_{t=0}+
\frac{\phi}{\lambda}\right)t+\left(|\tilde{x}(0)|+\frac{\phi}{\lambda^2}\right)
\label{eq:int2bounds}
\end{equation}

Furthermore, dividing (\ref{eq:int2bounds}) by $e^{\lambda t}$, it can be easily verified that for 
$t\to\infty$:

\begin{equation}
-\frac{\phi}{\lambda^2}\le\tilde{x}\le\frac{\phi}{\lambda^2}
\label{eq:txbounds}
\end{equation}

By imposing the bounds (\ref{eq:txbounds}) to (\ref{eq:int1bounds}), noting that $d(\tilde{x}e^{\lambda
t})/dt=\dot{\tilde{x}}e^{\lambda t}+\tilde{x}\lambda e^{\lambda t}$ and dividing again by $e^{\lambda
t}$, it follows that, for $t\to\infty$,
 
\begin{equation}
-2\frac{\phi}{\lambda}\le\dot{\tilde{x}}\le2\frac{\phi}{\lambda}
\label{eq:txdbounds}
\end{equation}

Finally, applying (\ref{eq:txbounds}) and (\ref{eq:txdbounds}) to (\ref{eq:sbounds}), one has

\begin{equation}
-6\phi\le\ddot{\tilde{x}}\le6\phi
\label{eq:txddbounds}
\end{equation}

In this way, the tracking error will be confined within the limits $|\tilde{x}|\le\phi/\lambda^2$,
$|\dot{\tilde{x}}|\le2\phi/\lambda$ and $|\ddot{\tilde{x}}|\le6\phi$. However, these bounds define 
a box that is not completely inside the boundary layer. Considering the demonstrated attractiveness 
and invariant properties of $S_\phi$, the region of convergence can be stated as the intersection 
of the boundary layer and the box defined by the preceding bounds. Therefore, it follows that the 
tracking error vector will exponentially converge to a closed region $\Phi=\{\mathbf{\tilde{x}}\in
\mathbb{R}^3\:|\:|s(\mathbf{\tilde{x}})|\le\phi\mbox{ and }|\tilde{x}|\le\phi/\lambda^2\mbox{ and }
|\dot{\tilde{x}}|\le2\phi/\lambda\mbox{ and }|\ddot{\tilde{x}}|\le6\phi\:\}$. It should be highlighted
that the convergence region $\Phi$ is in perfect accordance with the bounds proposed by \citet{ijac2009} 
for $n^\mathrm{th}$-order nonlinear systems subject to smooth sliding mode controllers.

Now, in order to obtain a good approximation to the disturbance $d$ and to enhance the tracking performance 
inside the convergence region $\Phi$, the estimate $\hat{d}$ will be computed directly by a neural network.
Due to its simplicity and fast convergence feature, radial basis functions (RBF) are used as activation 
functions and the related tracking error as input. In this case, the output of the network is defined as: 

\begin{equation}
\hat{d}(\mathbf{\tilde{x}})=\sum^M_{i=1}w_i\cdot\varphi_i(||\mathbf{\tilde{x}-t}||)
\label{eq:rbf}
\end{equation}

\noindent
where $\varphi_i(\cdot)$ are the activation functions and $\mathbf{t}$ a vector containing the coordinates of the center of each 
activation function.

Now, the signal $\nu=\tilde{x}^{(n)}+k_{n-1}\tilde{x}^{(n-1)}+\ldots+k_1\dot{\tilde{x}}+k_0\tilde{x}-\hat{d}$ is used to train 
the neural network and the weights of the output layer are adjusted using the pseudo-inverse matrix. 

Considering a training set $T=\{(\mathbf{\tilde{x}},d)_1,(\mathbf{\tilde{x}},d)_2,\ldots,(\mathbf{\tilde{x}},d)_p\}$ and

\begin{equation}
\left[\begin{array}{cccc}
\varphi_{11} &\varphi_{12} &\cdots &\varphi_{1M}\\
\varphi_{21} &\varphi_{22} &\cdots &\varphi_{2M}\\
\vdots       &\vdots       &\ddots &\vdots      \\
\varphi_{p2} &\varphi_{p2} &\cdots &\varphi_{pM}
\end{array} \right]
\left[\begin{array}{c}
w_1\\ w_2\\ \vdots\\ w_M
\end{array} \right]
=\left[\begin{array}{c}
d_1\\ d_2\\ \vdots\\ d_p
\end{array} \right]
\quad\therefore\quad
[\varphi]\{w\}=\{d\}
\label{eq:train}
\end{equation}

\noindent
the RBF weights are computed with the pseudo-inverse $[\varphi]^+$

\begin{equation}
\{w\}=[\varphi]^+\{d\}
\label{eq:weights}
\end{equation}

\noindent
and approximation error, $E$, by the euclidean norm

\begin{equation}
E=||\{d\}-[\varphi]\{w\}||
\label{eq:error}
\end{equation}

In the following section some numerical simulations are presented in order to evaluate the performance 
of the neural network based sliding mode controller.

\section*{SIMULATION RESULTS}

The simulation studies were performed with sampling rates of 500 Hz for control system and 1 kHz for dynamic 
model. The differential equations of the dynamic model were numerically solved with the fourth order Runge-Kutta 
method.

\begin{figure}[htb]
\centering
\mbox{
\subfigure[Tracking performance.]{\label{fig:graf1} 
\includegraphics[width=0.45\textwidth]{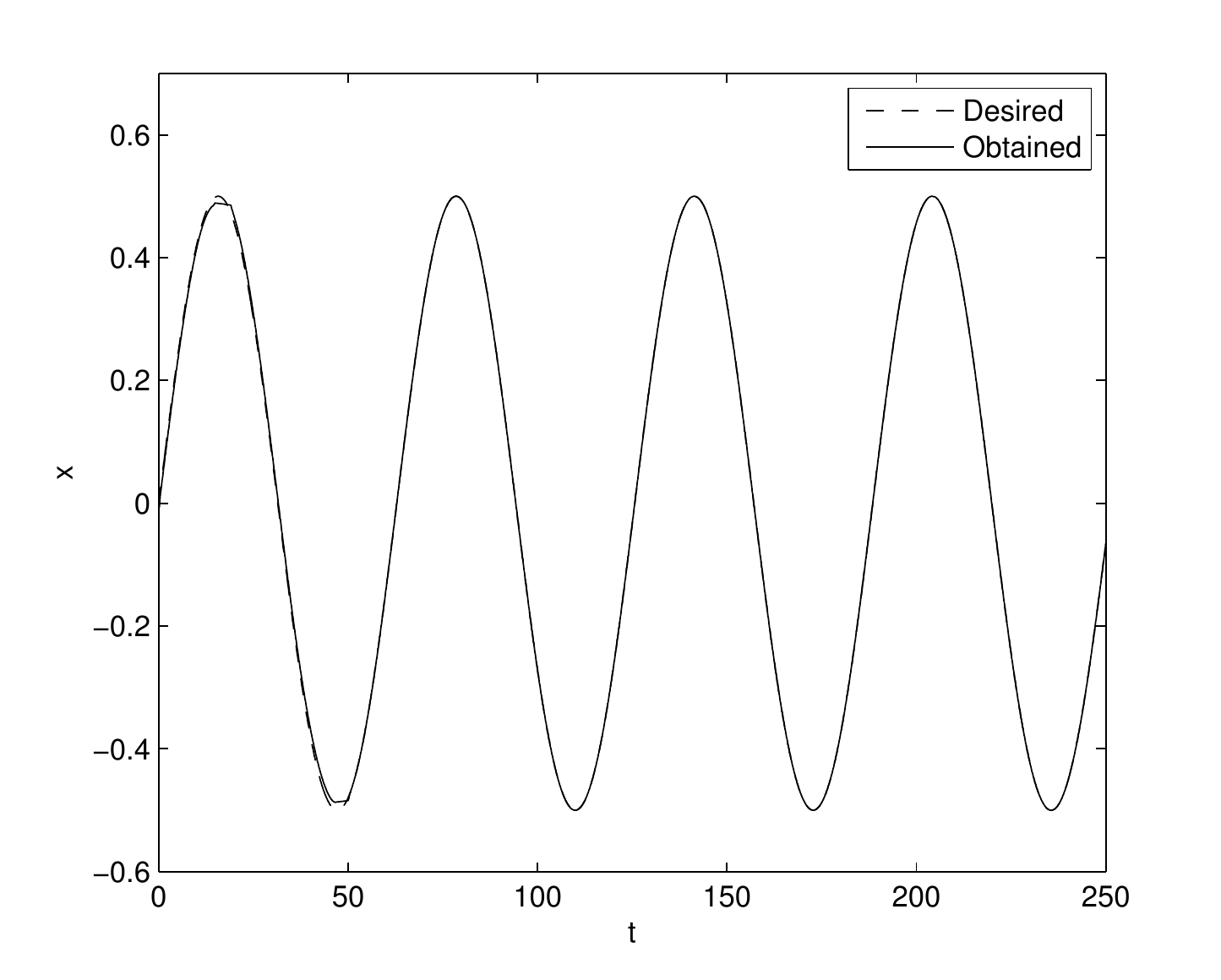}}
\subfigure[Control voltage.]{\label{fig:graf2} 
\includegraphics[width=0.45\textwidth]{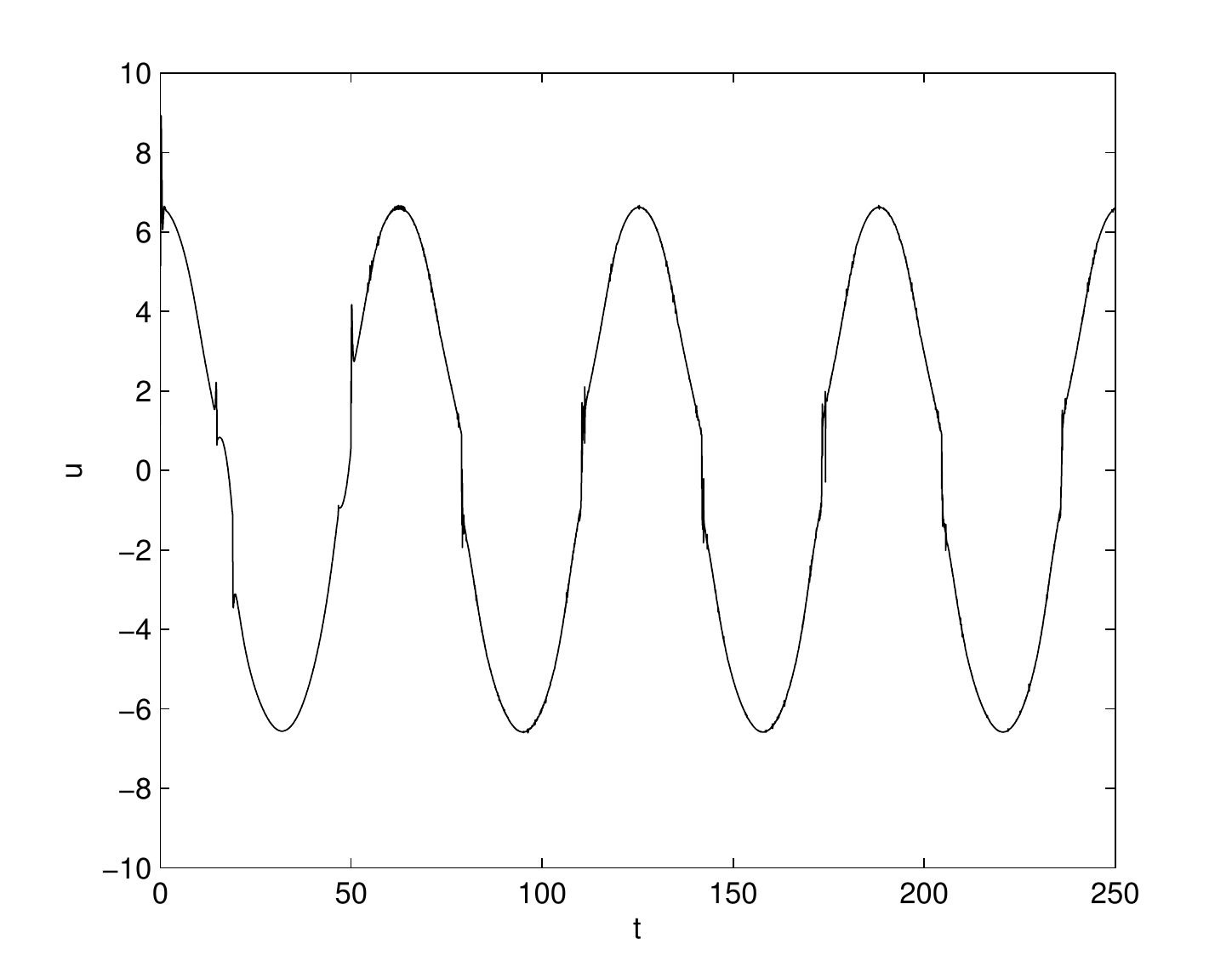}}
}
\caption{Tracking performance with $x_d=0.5\sin(0.1t)$ m.}
\label{fig:sim1}
\end{figure}

It was assumed in the simulation studies that the model parameters were not exactly known and nonlinear 
functions were considered for $g_l(u)=k_l(u+0.2\sin u-\delta_l)$ and $g_r(u)=k_r(u-0.2\cos u-\delta_r)$, 
with $k_l=k_r=2\times10^{-6}$ m/V. On this basis, considering a maximal uncertainty of $\pm10\%$ over 
the value of $k_v$ and variations of $\pm20\%$ in the supply pressure, $P_s=7[1+0.2\sin(x)]$ MPa, the 
estimates $\hat{k}_v=2\times10^{-6}$ m/V and $\hat{P}_s=7$ MPa were chosen for the computation of 
$\hat{b}$ in the control law. The other model and controller parameters were $\rho=850$ kg/m$^3$, 
$C_d=0.6$, $w=2.5\times10^{-2}$ m, $A_p=3\times10^{-4}$ m$^2$, $C_{tp}=2\times10^{-12}$ m$^3$/(s Pa), 
$\beta_e=700$ MPa, $V_t=6\times10^{-5}$ m$^3$, $M_t=250$ kg, $B_t=100$ Ns/m, $K_s=75$ N/m, $\delta_l=
-1.1$ V, $\delta_r=0.9$ V, $\lambda=8$, $\varphi=4$, $\gamma=1.2$, $\delta=1.1$, $\phi=1$ and $\eta=0.1$.
Figure~\ref{fig:sim1} shows the obtained results for the tracking of $x_d=0.5\sin(0.1t)$ m. It should be
highlighted that the first 50 seconds of each simulation study were used to train the neural network.

As observed in Fig.~\ref{fig:sim1}, even in the presence of uncertainties with respect to model 
parameters and an unknown dead-zone input, the neural network based sliding mode controller is 
able to provide trajectory tracking with no chattering at all. 

\begin{figure}[htb]
\centering
\mbox{
\subfigure[Without neural network compensation.]{\label{fig:graf3} 
\includegraphics[width=0.45\textwidth]{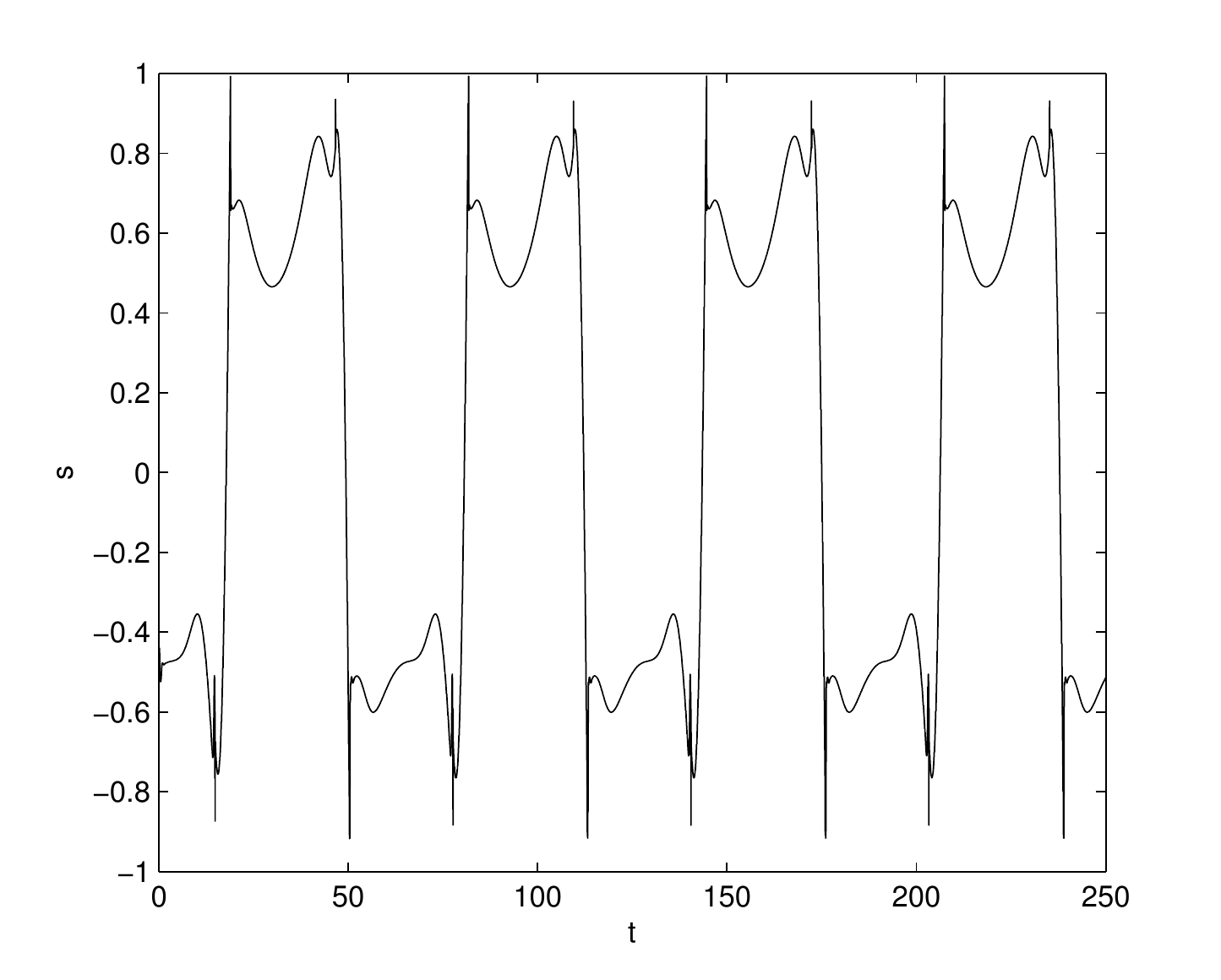}}
\subfigure[With neural network compensation.]{\label{fig:graf4} 
\includegraphics[width=0.45\textwidth]{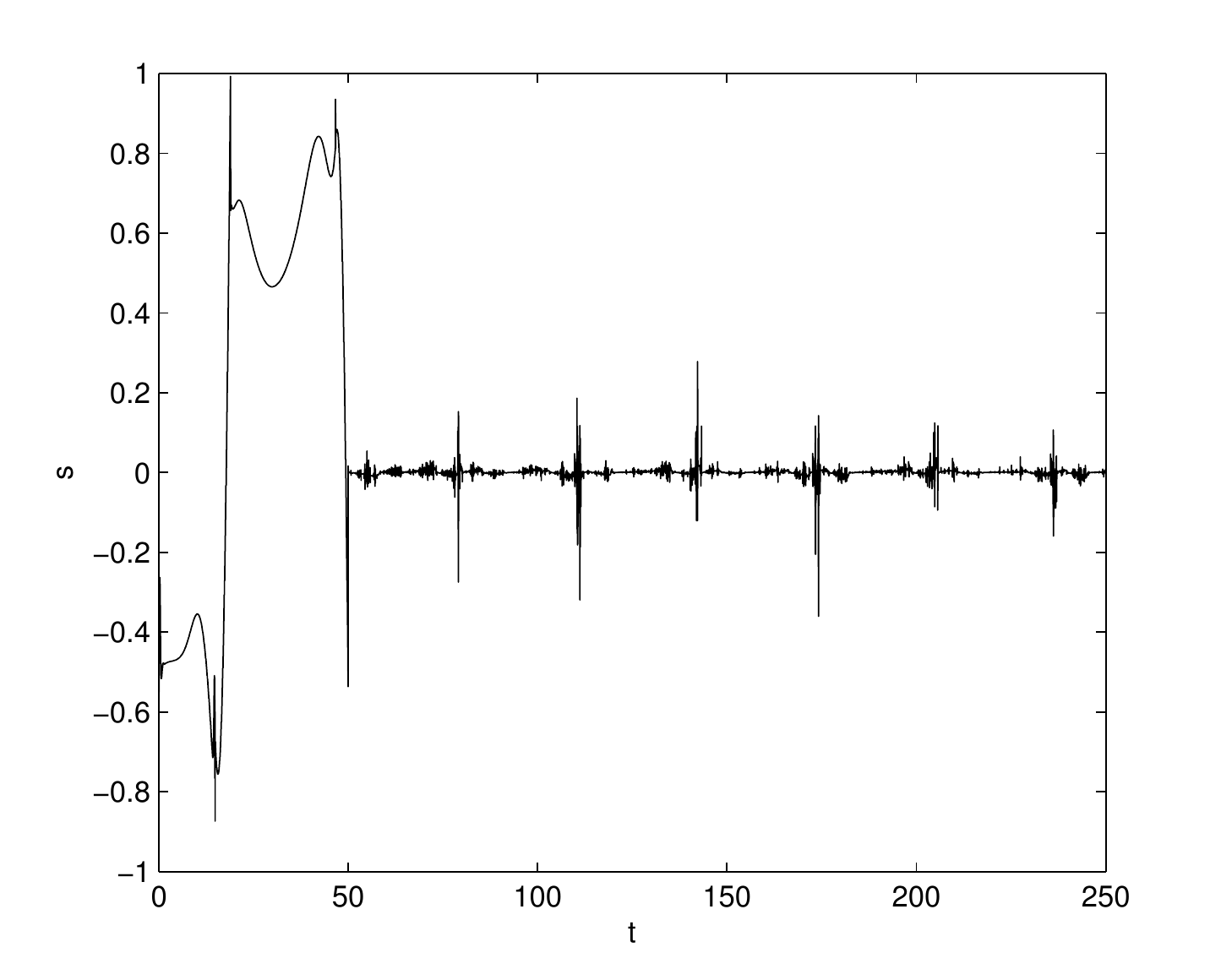}}
}
\caption{Comparison of the tracking error measure with and without neural network compensation.}
\label{fig:sim2}
\end{figure}

The improved performance of the proposed controller can be clearly ascertained if the evolution 
in time of the related tracking error measure, $s$, is compared with the result obtained with 
a conventional smooth sliding mode controller. For simulation purposes, the proposed controller
can be easily converted to the classical one by setting $\hat{d}=0$. Figure~\ref{fig:sim2} shows 
the obtained results. It can be easily verified in Fig~\ref{fig:graf4} that after 50 seconds,
when the compensation scheme is enabled, the magnitude of the tracking error measure $s$ is 
significantly reduced.

\section*{CONCLUDING REMARKS}

The present work addressed the problem of controlling electro-hydraulic systems subject to an unknown
dead-zone. A neural network based sliding mode controller was implemented to deal with the trajectory 
tracking problem. The boundedness and convergence properties of the closed-loop systems was proven 
using Lyapunov stability theory. The control system performance was also confirmed by means of 
numerical simulations. The neural network algorithm could automatically recognize the dead-zone 
nonlinearity and previously compensate for its undesirable effects. 

\section*{ACKNOWLEDGMENTS}

The authors would like to acknowledge the support of the Brazilian National Research Council (CNPq), the Brazilian Coordination for 
the Improvement of Higher Education Personnel (CAPES) and the German Academic Exchange Service (DAAD).

\end{document}